\documentclass[pra,twocolumn,showpacs,eqsecnum,superscriptaddress]{revtex4}
\usepackage{makeidx}
\usepackage{eurosym}
\usepackage{amssymb}
\usepackage{amsfonts}
\usepackage{amsmath}
\usepackage{setspace}
\usepackage[english]{babel}
\usepackage{graphicx}
\usepackage[applemac]{inputenc}
\usepackage{color}
\usepackage{ifthen}
\usepackage{float}
\usepackage{verbatim}
\usepackage{subfigure}
\begin{document}

\title{Dynamical Phase Transition of two-component Bose-Einstein condensate with nonlinear tunneling in an optomechanical cavity-mediated double-well system}
\author{Qing Li}
\affiliation{Institute of Theoretical Physics, Lanzhou University, Lanzhou $730000$, China}
\author{Lei Tan}
\email{tanlei@lzu.edu.cn}
\affiliation{Institute of Theoretical Physics, Lanzhou University, Lanzhou $730000$, China}
\author{Jin-Lou Ma}
\affiliation{Institute of Theoretical Physics, Lanzhou University, Lanzhou $730000$, China}
\author{Huai-Qiang Gu}
\affiliation{School of Nuclear Science and Technology, Lanzhou University, Lanzhou $730000$, China}
\author{Yun-Xia Shi}
\affiliation{Institute of Theoretical Physics, Lanzhou University, Lanzhou $730000$, China}
\author{Wu-Ming Liu}
\affiliation{Beijing National Laboratory for Condensed Matter Physics, Institute of Physics, Chinese Academy of Sciences, Beijing 100190, China}

\date{\today}
\begin{abstract}
We investigate the dynamical phase transition of two-component Bose-Einstein condensate with nonlinear tunneling, which is trapped inside a double-well and dispersively coupled to a single mode of a high-finesse optical cavity with one moving end mirror driven by a single mode standing field. The nonlinear tunneling interaction leads to an increase of stability points and riches the phase diagram of the system. It is shown that the appearance of the moving end mirror speeds up the tunneling of Bose-Einstein condensates, which makes population difference between two wells and regulates the number of the stability points of the system.

\end{abstract}


\maketitle

\section{Introduction}
The system of Bose-Einstein condensates (\textbf{BECs}) in double-well (DW) potentials is an important platform for quantum manipulation due to its highly controllable experimental parameters \cite{Milburn,Albiez}, which has the potential to demonstrate a wide range of fundamental quantum phenomena with regard to manipulating the tunneling dynamics governed by the two-body interactions locally and single particle tunneling strength between wells. Some exciting rich phasespace dynamics in theoretical and experimental studies  have been revealed. These include the dynamics of spin-orbit-coupled condensates \cite{Zhang2012Josephson,Garcia2014Josephson,Niu2016Tunnelling}, the existence of nonlinear steady state\cite{holthaus2001towards}, the cross structure of the level\cite{witthaut2006towards,wu2006commutability}, the nonlinear Landau-Zener tunneling\cite{liu2002theory}, and the the nonlinear Josephson oscillation\cite{smerzi1997quantum,milburn1997quantum}. Most recently, the tunneling probabilities of few bosons\cite{Dutta}, the nonequilibrium dynamical ion transfer\cite{Klumpp}, the asymmetric many-body loss\cite{Denis}, the dynamical phase transition of binary species BECs\cite{Tian}, Interaction blockade\cite{Cosme}and the interaction-modulated tunneling dynamics\cite{Maraj} have also been explored, respectively.

To obtain strong \textbf{atom}-field nonlinearity and tailor the atom-field coupling effectively, a great step was made as two groups succeeded independently in coupling a BECs to a single-cavity mode in experiment\cite{Brennecke,Colombe}. In this sense, BEC DW systems
with one or two wells coupled to the cavity fields have been discussed previously. Homodyne measurements\cite{Corney}, the interplay dynamics of\cite{Zhang}, the mean-field dynamics of a Bose Josephson junction\cite{Zhang1,tanlei}, the outcomes of the atom-field nonlinearity\cite{Larson} are investigated.  Nondemolition measurements have also been proposed based on this system\cite{Zuppardo}. In recent years, optomechanical cavities have emerged and become another ideal and irreplaceable system to study the strong matter-field interaction\cite{ensemble}. Such a system demonstrates the interaction between the movable oscillator and the cavity field via the radiation pressure and becomes a new platform for the study of  ground-state cooling of the vibrational modes of a mechanical oscillator\cite{Yasir}, coherent quantum noise cancellation\cite{Motazedifard}, the steady-state bipartite entanglement and quadrature squeezing\cite{Dalafi}, Bistability\cite{Yasir2014}, electromagnetically induced transparency (EIT) and Fano Resonances\cite{Yasir2015}, the laser phase noise \cite{Dalafi94}, and the emergences of the entanglement\cite{Chiara,Paternostro}.
Given the wealth of effects resulting from the hybrid system of BECs in an optomecanical cavity, it is natural to ask
for the new phenomena stemming from an optomechanical cavity-mediated BECs DW system.

Note that, the nonlinear tunneling can be omitted as it is several orders of magnitude smaller than the linear tunneling strength in the weak interaction range. Actually, new phenomena will occur when one varies the interaction strength from the weak to the strong limit. The correlated tunneling was firstly observed in a sample of rubidium atoms in the few-atom and strong coupling limit\cite{Folling}. It was shown that the two atoms evolves from Rabi osicllations to correlated pair tunneling with the increase of the interaction strength\cite{Zollner}. Following from this finding, there has been a great deal of efforts devoted to the Bose-Hubbard model with nonlinear tunneling, such as the atom-pair tunneling and quantum phase transition in the strong-interaction regime\cite{Liang}, the fragmented condensate\cite{Zhu}. Very recently, the quantum phase transitions between a Josephson phase, a self-trapping phase, and a phase-locking are found in an extended two-mode Bose-Hubbard model with nonlinear tunneling\cite{Rubeni}. This raises the prospect of investigations into nonlinear tunneling effects in BECs DW beyond the previous work mentioned above.

Motivated by the above prospects, we investigated the mean-field dynamics of a two-component BECs DW  with the nonlinear tunneling, which are trapped in a high-finesse optical cavity with a moving end mirror due to the reasons that much more complicated and achievable states can be obtained due to the interplay
of intra-species and inter-species interaction of different species BECs. We find that the nonlinear tunneling increases the stability points and enriches the phase diagram of the system. Furthermore, the coupling strength between the cavity and the moving end mirror and the detuning between the pump field and the moving end mirror can regulate the number of the stability points of the system, and then control the dynamics of the system.

The paper is organized as follows: the Hamiltonian of the system and the dynamical equation are presented in Sec.\uppercase\expandafter{\romannumeral2}. In Sec. \uppercase\expandafter{\romannumeral3}, we derive the classical model of the system using the mean-field theory. Sec.\uppercase\expandafter{\romannumeral5} is devoted to discuss the stationary points and energy contours of the BECs DW. Finally, the conclusion is summarized in Sect. \uppercase\expandafter{\romannumeral6}.

\section{system Hamiltonian and the dynamical equation}
\begin{figure}[H]
\centering
\includegraphics[width=3in]{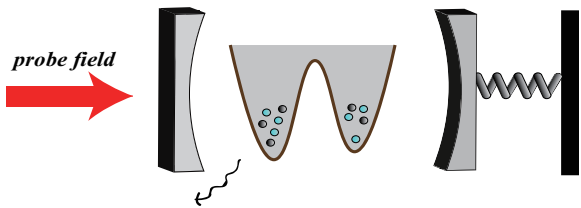}
\caption{The double-well trap for two-component Bose-Einstein condensates with $N_1({N_1}^{\prime})$ and $N_2({N_2}^{\prime})$ the number of particles in the cavity with the mirror. The number $1$ and $2$ represent the left well and the right one, respectively.}
\end{figure}
Consider an optomechanical cavity containing a two-component BECs DW with a fixed mirror and a movable mirror of mechanical frequency $\omega_{m}$, which is driven by a pump field with frequency $\omega_{p}$,  as schematically shown in Fig. $1$.  The numbers of each component is $N$.  The
Hamiltonian of the system can be written as
\begin{equation}\label{}
H=H_a+H_F+H_M+H_{couple},
\end{equation}
where $H_{a}$ describes the behavior of the atomic modes (BECs) and their interactions with each other, $H_{F}$ gives the energy of the single mode cavity, $H_{M}$ is related to the mechanical resonator and its association with the pump field, $H_{couple}$ accounts for the interaction of the single mode cavity with the mechanical resonator and the atoms. The atom-pair tunneling term (nonlinear tunneling) is also included in this Hamiltonian\cite{Rubeni}. In the two-mode approximation\cite{milburn1997quantum}, the canonical BECs DW  Hamiltonian reads (assuming $\hbar=1$)
\begin{eqnarray}\label{}
H_a&=&-\Omega_1(b_{_{1}}^{\dagger} b_{_{2}}+b_{_{2}}^{\dagger} b_{_{1}})-\Omega_2(c_{_{1}}^{\dagger} c_{_{2}}+c_{_{2}}^{\dagger} c_{_{1}})\nonumber\\&+&
\frac{V_1}{2}(b_{_{1}}^{\dagger} b_{_{1}}^{\dagger} b_{_{1}}b_{_{1}}+b_{_{2}}^{\dagger}b_{_{2}}^{\dagger}b_{_{2}}b_{_{2}})\nonumber\\&+&
\frac{V_2}{2}(c_{_{1}}^{\dagger} c_{_{1}}^{\dagger} c_{_{1}}c_{_{1}}+c_{_{2}}^{\dagger}c_{_{2}}^{\dagger}c_{_{2}}c_{_{2}})\nonumber\\&+&
\frac{{V_1}'}{2}b_{_{1}}^{\dagger}b_{_{1}}c_{_{1}}^{\dagger} c_{_{1}}+\frac{{V_2}'}{2}b_{_{2}}^{\dagger}b_{_{2}}c_{_{2}}^{\dagger}c_{_{2}}\nonumber\\&-&
\frac{S_{1}}{2}(b_{_{1}}^{\dagger} b_{_{1}}^{\dagger} b_{2} b_{2}+b_{_{2}}^{\dagger} b_{_{2}}^{\dagger} b_{1} b_{1})\nonumber\\&-&
\frac{S_{2}}{2}(c_{_{1}}^{\dagger} c_{_{1}}^{\dagger} c_{2} c_{2}+c_{_{2}}^{\dagger} c_{_{2}}^{\dagger} c_{1} c_{1}),
\end{eqnarray}
For simplicity, we assume that $\Omega_1=\Omega_2=\Omega$, $V_1=V_2=V$, $V_1'=V_2'=V'$ \cite{smerzi1997quantum}, $S_1=S_2=S$, the subscripts $1$ and $2$ represent the localised modes in the left and right potential wells, respectively. $b_{_{i}}^{\dagger}(b_{_{i}})$ and $c_{_{i}}^{\dagger}(c_{_{i}})$ are the creation (annihilation) operators of the two atoms modes, respectively. $\Omega$ means the parameter of tunneling between the two modes. The parameter $V$($V'$) denotes the interaction between atoms of the homogeneous(heterogeneous) species. $S$ is the coupling strength for the atom-pair tunneling. The two-mode model assumes two stationary wave functions is such that the two lowest states are closely spaced and well separated from higher levels of the potential, and that many-particle interactions do not significantly change double well\cite{milburn1997quantum}. Correspondingly, the Hamiltonian of the single-mode optical field is
\begin{equation}\label{}
H_F=\omega_{c}a^{\dagger}a+\eta(t)e^{-i\omega_pt}a^{\dagger}+\eta(t)^{*}e^{i\omega_pt}a,
\end{equation}
where $\omega_{c}$ and $\omega_{p}$ are the cavity and pump field frequency, respectively. $\eta(t)$ represents the optical amplitude of the pump, here, we assume that the amplitude of the pump varies slowly, that is $|\dot{\eta}/ \eta|\ll\omega_p$\cite{zhang2008cavity}. The Hamiltonian of the moving end mirror $H_M$ can be read as\cite{ghobadi2013towards}
\begin{equation}\label{}
H_M=\omega_m d^{\dagger}d,
\end{equation}
where $\omega_m$ is the frequency of the moving end mirror. In the two-mode approximation, due to the coupling strength between the cavity mode and the atomic tunneling is much smaller than the overlaps between the atomic modes and the cavity mode, that is $J_{12}({J_{12}}^{\prime})\ll J_{1,2}({J_{1,2}}^{\prime})$. Therefore, we have dropped some terms $U_0a^{\dagger}aJ_{12}({b_1}^{\dagger}b_2+{b_2}^{\dagger}b_1)$ and $U_0a^{\dagger}a{J_{12}}^{\prime}({c_1}^{\dagger}c_2+{c_2}^{\dagger}c_1)$\cite{zhang2008mean}. Now, we write the Hamiltonian governed the field-condensate and the moving end mirror interaction as\cite{maschler2005cold,ghobadi2013towards}
\begin{eqnarray}\label{}
H_{couple}&=&U_0 a^{\dagger}a(J_1n_1+J_2n_2+{J_1}^{\prime}{n_1}^{\prime}+{J_2}^{\prime}{n_2}^{\prime})\nonumber\\&&
-G_0(a^{\dagger}+a)(d^{\dagger}+d),
\end{eqnarray}
where $U_0={g_0}^2/(\omega_c-\omega_a)$ is the light shift per photon, $g_0$ being the atom-field coupling strength at an antinode. $G_0=\frac{\omega_c}{L}\sqrt{\frac{\hbar}{2M\omega_m}}$ is the coupling strength between the cavity and the moving end mirror, $\triangle_a=\omega_c-\omega_a$ is the far-off detuning between the atoms and field frequency\cite{ayub2014dynamical,ghobadi2013towards}. $J_1({J_1}^{\prime})$ and $J_2({J_2}^{\prime})$ account for the overlapping between the atomic modes and the cavity mode\cite{zhang2008cavity}. In this section, we do not discuss the case that $J_1({J_1}^{\prime})=J_2({J_2}^{\prime})$ because it illustrates the atoms do not interact with the single mode field. In other words, the cavity can not influence the distribution of the atoms in double well. Therefore, we focus on another case, $J_1({J_1}^{\prime})\not=J_2({J_2}^{\prime})$. In the rotating-wave approximation, the total Hamiltonian leads to coupled quantum Langevin equations for the annihilation operators of BEC, the cavity and the moving end mirror, viz.,
\begin{widetext}
\begin{eqnarray}\label{}
i\dot{b_1}&=&-\Omega b_2+V{b_1}^{\dagger}b_1b_1+\frac{V^{\prime}}{2}{c_1}^{\dagger}c_1b_1+J_1U_0a^{\dagger}ab_1-S{b_1}^{\dagger}b_2b_2\nonumber\\
i\dot{b_2}&=&-\Omega b_1+V{b_2}^{\dagger}b_2b_2+\frac{V^{\prime}}{2}{c_2}^{\dagger}c_2b_2+J_2U_0a^{\dagger}ab_2-S{b_2}^{\dagger}b_1b_1\nonumber\\
i\dot{c_1}&=&-\Omega c_2+V{c_1}^{\dagger}c_1c_1+\frac{V^{\prime}}{2}{b_1}^{\dagger}b_1c_1+{J_1}^{\prime}U_0a^{\dagger}ac_1-S{c_1}^{\dagger}c_2c_2\nonumber\\
i\dot{c_2}&=&-\Omega c_1+V{c_2}^{\dagger}c_2c_2+\frac{V^{\prime}}{2}{b_2}^{\dagger}b_2c_2+{J_2}^{\prime}U_0a^{\dagger}ac_2-S{c_2}^{\dagger}c_1c_1\nonumber\\
i\dot{a}&=&[\omega_c+U_0(J_1n_1+J_2n_2+{J_1}^{\prime}{n_1}^{\prime}+{J_2}^{\prime}{n_2}^{\prime})]a-G_0d\nonumber-i\kappa a+\eta(t)e^{-i\omega_pt}\nonumber\\
i\dot{d}&=&\omega_m d-G_0 a
\end{eqnarray}
\end{widetext}
The parameter $-i\kappa a$ in Eq. $(2.6)$ represents the dissipation of the cavity and $\kappa$ is the dissipation rate correspondingly. From Eq. $(2.6)$, it's not difficult to find out that the change of atomic number has a great relationship with the photon number in the cavity field through $J_iU_0a^{\dagger}ab_i$ and ${J_i}^{\prime}U_0a^{\dagger}ac_i$ terms. In other words, the BJJ is tilted with the photon distribution which depends on the atom number. Therefore, we can use the varying pump frequency to regulate the property of the BJJ.

\section{THE MODEL OF THE SYSTEM}
Under the mean-field approximation, we consider atomic and photonic operators to be classical quantities, namely $b_1=\sqrt{N_1}e^{i\theta_1}$, $b_2=\sqrt{N_2}e^{i\theta_2}$, $c_1=\sqrt{{N_1}^{\prime}}e^{i{\theta_1}^{\prime}}$, $c_2=\sqrt{{N_2}^{\prime}}e^{i{\theta_2}^{\prime}}$, $a=\alpha$, $d=\beta$.
$\theta_{1,2}$ and ${\theta_{1,2}}^{\prime}$ describe the corresponding phase of the atom. Moreover, $N_{1,2}$ and ${N_{1,2}}^{\prime}$ are the total atomic numbers of $b$ and $c$ in two wells, respectively. In this model, we assume that the total atomic numbers of $b$ and $c$ are equal, $i_\cdot e_\cdot$, $N_b=N_c=N$\cite{ng2003quantum}.

In this system, it is clear from Eq. $(2.6)$ that the relaxation time scale of the cavity mode is of the order of $1/\kappa$, $\kappa\sim2\pi\times10^6$ Hz, which is much shorter than the oscillation period of a bare BECs DW \cite{goldstein1998dressed}, which is of the order of $1/\Omega$, usually, $\Omega$ is of the order of $2\pi\times10^{1-2}$ Hz in the real experiment\cite{albiez2005direct}. The end mirror of the cavity with the frequency $\omega_m$ is of the order of $2\pi\times10^5$ Hz\cite{yasir2014exponential}. This implies that the cavity field follows the motion of the condensates adiabatically\cite{horak2000coherent}. Thus, from Eq. $(2.6)$, one reads
\begin{widetext}
\begin{eqnarray}
\langle a \rangle=\alpha(t)=\frac{\eta(t)e^{-i\omega_pt}(\omega_m-\omega_p)}{{G_0}^2-[\omega_c+U_0(J_1N_1+J_2N_2+{J_1}^{\prime}{N_1}^{\prime}+{J_2}^{\prime}{N_2}^{\prime})-i\kappa-\omega_p](\omega_m-\omega_p)},
\end{eqnarray}
and the mean photon number is
\begin{align}
\langle a^{\dagger}a\rangle&=|\alpha(t)|^2{}\nonumber\\&{}=
\frac{{\eta(t)}^2{{\Delta}^{\prime}}^2}{{G_0}^4-2{G_0}^2{\Delta}^{\prime}[\Delta-\delta U_0\frac{(N_1-N_2)}{2}-{\delta}^{\prime}{U_0}^{\prime}\frac{{(N_1}^{\prime}-{N_2}^{\prime})}{2}]+{{\Delta}^{\prime}}^2{[(\Delta-\delta U_0\frac{(N_1-N_2)}{2}-{\delta}^{\prime}{U_0}^{\prime}\frac{{(N_1}^{\prime}-{N_2}^{\prime})}{2})}^2+{\kappa}^2]},\label{two}
\end{align}
here $\Delta=\omega_p-\omega_c-(J_1+J_2)\frac{NU_0}{2}+({J_1}^{\prime}+{J_2}^{\prime})\frac{NU_0}{2}$, ${\Delta}^{\prime}=\omega_p-\omega_m$, $\delta=J_1-J_2$, ${\delta}^{\prime}={J_1}^{\prime}-{J_2}^{\prime}$;
$\delta$ and ${\delta}^{\prime}$($\delta$=${\delta}^{\prime}$) represent the coupling difference between the two atomic modes to the double well.

Introducing the dimensionless parameters $Z_b=\frac{N_1-N_2}{N}$ and $Z_c=\frac{{N_1}^{\prime}-{N_2}^{\prime}}{N}$, which describe the population difference of the two modes atoms between the double well. Therefore, we rewrite Eq. (3.2) as
\begin{eqnarray}
\alpha(Z_b,Z_c,t)^2=\frac{A(t)^2 E^2}{D^4-2D^2E(Z_b+Z_c-B)+E^2[(Z_b+Z_c-B)^2+C^2]},\label{three}
\end{eqnarray}
\end{widetext}
where $A(t)={\eta(t)}/[\frac{\delta NU_0}{2}]$, $B={\Delta}/[\frac{\delta NU_0}{2}]$, $C={\kappa}/[\frac{\delta NU_0}{2}]$, $D=G_0/[\frac{\delta NU_0}{2}]$, $E={\Delta}^{\prime}/[\frac{\delta NU_0}{2}]$. We regard $A(t)$, $B$, $C$ as the reduced pumping strength, reduced detuning and reduced loss rate, respectively. And we may understand $D$ as the reduced coupling strength between the cavity and the moving end mirror. It can be found from Eq. $(3.3)$ that, the mean photon number is a Lorentzian at $z_b+z_c=B+D^2/E$ with a width $2C$, which is a function of $z_b$ and $z_c$. This is because atoms and mirror are forced to vibrate due to the cavity mode. The addition of the mirror makes the peak position of the photon number distribution move to the left by $D^2/E$.

Substituting Eq. $(3.3)$ into Eq. $(2.6)$ and defining the relative phases of the atoms as $\phi_b=\theta_1-\theta_2$ and $\phi_c={\theta_1}^{\prime}-{\theta_2}^{\prime}$, the Eq. $(2.6)$ can be rewritten in terms of $z_b(z_c)$ and the phase difference $\phi_b(\phi_c)$ as
\begin{equation}
\dot{\phi_{b}}=\frac{z_{b}}{\sqrt{1-z_{b}^{2}}}\cos\phi_{b}+r_{b}z_{b}+\frac{r_{bc}z_{c}}{2}+\Lambda z_{b}\cos(2\phi_{b})+\frac{\delta
U_{0}}{2\Omega}|\alpha|^{2}\nonumber,
 \label{}
\end{equation}
\begin{equation}
\dot{z_{b}}=-\sqrt{1-z_{b}^{2}}\sin\phi_{b}-\Lambda (1-z_{b}^{2})\sin(2\phi_{b})\nonumber,
 \label{}
\end{equation}
\begin{equation}
\dot{\phi_{c}}=\frac{z_{c}}{\sqrt{1-z_{c}^{2}}}\cos\phi_{c}+r_{c}z_{c}+\frac{r_{bc}z_{b}}{2}+\Lambda z_{c}\cos(2\phi_{c})+\frac{\delta
U_{0}}{2\Omega}|\alpha|^{2}\nonumber,
 \label{}
\end{equation}
\begin{equation}
\dot{z_{c}}=-\sqrt{1-z_{c}^{2}}\sin\phi_{c}-\Lambda (1-z_{c}^{2})\sin(2\phi_{c}),
 \label{}
\end{equation}
where the time has been rescaled in units of the Rabi oscillation time $1/(2\Omega)$, $2\Omega t\rightarrow t$. $r_b$, $r_c$, $r_{bc}$ and $\Lambda$ are the interaction strengths against the tunneling strength, $r_b$=$r_c$=$r$=$NV/{2\Omega}$ express the interaction strength between homologous atoms and $r\geq r_{bc}$. We further define a Hamiltonian as a function of two conjugate variables $z_n(n=b,c)$ and $\phi_n(n=b,c)$. i.e., $\dot{z_n}=-\frac{\partial{H_n}}{\partial\phi_n}$, $\dot{\phi_n}=\frac{\partial{H_n}}{\partial z_n}$, therefore a Hamiltonian is
\begin{eqnarray}\label{}
H_b(z_b,{\phi}_b,t)&=&-\sqrt{1-z_{b}^{2}}\cos\phi_{b}+\frac{{r_{b}z_{b}}^2}{2}+\frac{r_{bc}z_{b}z_{c}}{2}-\frac{\Lambda {z_b}^2}{2}\nonumber\\&&
-\Lambda (1-z_{b}^{2})\cos^2\phi_{b}+\frac{\delta{U_0}}{2\Omega}F(z_b,z_c,t)\nonumber,
\end{eqnarray}
\begin{eqnarray}
H_c(z_c,{\phi}_c,t)&=&-\sqrt{1-z_{c}^{2}}\cos\phi_{c}+\frac{{r_{c}z_{c}}^2}{2}+\frac{r_{bc}z_{b}z_{c}}{2}-\frac{\Lambda {z_c}^2}{2}\nonumber\\&&
-\Lambda (1-z_{c}^{2})\cos^2\phi_{c}+\frac{\delta{U_0}}{2\Omega}F(z_b,z_c,t)\nonumber\\,
\end{eqnarray}
with
\begin{equation}\label{}
F(z_b,z_c,t)=\frac{{A(t)}^2}{C}\arctan\frac{-D^2+E(z_b+z_c-B)}{CE},
\end{equation}
The first five terms of the Hamiltonian in Eq. $(3.5)$ are the Hamiltonian of a bare BECs DW, Among them the first three terms  describe the energy cost due to the phase twisting between the two condensates, the interaction between atoms of the homogeneous species and the interaction between atoms of the heterogeneous species, respectively. Compared with \cite{tanlei}, the added fourth term and fifth terms of each Hamiltonian indicate the atom-pair tunneling in a double-well potential\cite{Rubeni}, which are caused by the nonlinear tunneling coupling and the phase twisting between the two condensates. In the standard Bose-Hubbard model, the nonlinear tunneling term is neglected, as they are small compared with the hopping energy and the on-site interaction.
However, the Eq. $(3.5)$ reveals that the nonlinear tunneling can change the distribution of energy contour of system and influence the dynamical of the atoms. And the last term of the two Hamiltonian are regarded as the cavity-field-induced tilt\cite{zhang2008mean}. If the pump strength changes with time, the Hamiltonian can be made explicitly time dependent. However, in this work, I focus on the case that the pump strength is a constant, viz, $\eta(t)\equiv\eta$. So, the Hamiltonian is conserved in time. In a simple mechanical analogy, $H_b(H_c)$ describes a nonrigid pendulum in with a tilted angle $\phi_b(\phi_c)$ and length proportional to $\sqrt{1-z_{b}^{2}}(\sqrt{1-z_{c}^{2}})$, which decreases with the angular momentum $z_b(z_c)$. But in this paper, $H_b(H_c)$ describes the stack of two nonrigid pendulum, the above description is one of them and another is tilt angle $2\phi_b(2\phi_c)$ and length proportional to  $\frac{\Lambda(1-z_{b}^{2})}{2}(\frac{\Lambda(1-z_{c}^{2})}{2})$.

As we all know, the energy of the system can be obtained for the conservative system. It has been demonstrated that the eigenstates of the system are related to the stationary points of phase-space level curves.
Next, we will explore the dynamics of a BECs DW  in the perspective of the phase-space level curves. First of all, we need to figure out the stationary points of the system by the equations $\frac{\partial{H_n}}{\partial z_n}=0, \frac{\partial{H_n}}{\partial\phi_n}=0$ $(n=b,c)$. The second equation suggests that $\phi=0$ or $\phi=\pi$, then, we can get the following expression from the first equation
\begin{widetext}
\begin{subequations}
\begin{equation}\label{}
f_1(z_b)=\frac{z_{b}}{\sqrt{1-z_{b}^{2}}}+r_{b}z_{b}+\frac{r_{bc}z_{c}}{2}+\Lambda z_{b}+\frac{\widetilde{A} E^2}{D^4-2D^2E(Z_b+Z_c-B)+E^2[(Z_b+Z_c-B)^2+C^2]}=0,
\end{equation}
\begin{equation}
f_1(z_c)=\frac{z_{c}}{\sqrt{1-z_{c}^{2}}}+r_{c}z_{b}+\frac{r_{bc}z_{b}}{2}+\Lambda z_{c}+\frac{\widetilde{A} E^2}{D^4-2D^2E(Z_b+Z_c-B)+E^2[(Z_b+Z_c-B)^2+C^2]}=0,
\end{equation}
\begin{equation}\label{}
f_2(z_b)=-\frac{z_{b}}{\sqrt{1-z_{b}^{2}}}+r_{b}z_{b}+\frac{r_{bc}z_{c}}{2}+\Lambda z_{b}+\frac{\widetilde{A} E^2}{D^4-2D^2E(Z_b+Z_c-B)+E^2[(Z_b+Z_c-B)^2+C^2]}=0,
\end{equation}
\begin{equation}
f_2(z_c)=-\frac{z_{c}}{\sqrt{1-z_{c}^{2}}}+r_{c}z_{b}+\frac{r_{bc}z_{b}}{2}+\Lambda z_{c}+\frac{\widetilde{A} E^2}{D^4-2D^2E(Z_b+Z_c-B)+E^2[(Z_b+Z_c-B)^2+C^2]}=0,
\end{equation}
\end{subequations}
\end{widetext}
here ${\widetilde{A}=\delta U_0{A(t)}^2}/{2\Omega}$. Another stationary point of the system is worked by $\frac{\partial{H_n}}{\partial z_n}=0,\frac{\partial{H_n}}{\partial\phi_n}=0$ $(n=b,c)$ when $\phi\not=0$ and $\phi \not=\pi$, which will be discussed using the  numerical method. For simplicity, we focus on  $z_b=z_c=z$, $\phi_b=\phi_c=\phi$ and analyze the character (minimum, saddle, or
maximum) of the possible stationary points by the corresponding Hessian matrices, which is the square matrix of second-order partial derivatives of the Hamiltonian about $z_n$ and $\phi_n$.

\section{Stationary points and Energy contours of the Bose Josephson junction}
\textit{Without the moving end mirror}. In order to focus on the impact of the nonlinear tunneling on the system dynamics, we study the solutions of the energy contours Eq. $(3.5)$ and the stationary point Eq. $(3.7)$ with $D=0$, $E=\omega_{p}$. As can be seen from Eqs. $(3.5)$ and $(3.7)$, the nonlinear tunneling contributes a term linear in $z^2$ to the Hamiltonina $H_b$ $(H_c)$ and, in turn, a term linear in $z$ to the functions $f_{1,2}$. It is natural to expect that this term can arise new roots and make the phase portrait of the system may be quantitatively or even qualitatively different from that of without the nonlinear tunneling. For the interplay between the cavity and the BECs, the roots of Eq. $(3.7)$  have to be solved numerically. When the nonlinear tunneling strength is small, the corresponding phase diagram is similar to the one without nonlinear tunneling term\cite{tanlei}, as shown in Fig. $2(a)$. There are three stationary points (two minimum and a saddle point) along the line $\phi=0$, and five stationary points (three maximum and two saddle points) along the line $\phi=\pi$. For the strong pair tunneling case, the phase diagram of the system has undergone tremendous changes, as illustrated in Fig. $2(b)$, and the corresponding stationary points of the map are given in Figs. $2(c)$-$2(e)$.  At first, a stable point (a maximum) appears along near the line $\phi=\frac{\pi}{2}$, its specific location can be seen in Fig. $2(d)$ (the stability point appears along near by $\pi/2$ from Fig. $2(b)$). Secondly, the system has experienced a transition from oscillating-phase-type self-trapping\cite{Wang2006Periodic} to running-phase-type self-trapping\cite{Wang2006Periodic} near by $z=-1$ and $z=1$ along the lines $\phi=\pi$, and the stationary points have a transition from two saddle points to two minimum and from one maximun to saddle points.

In Fig. $3$, we plot the time evolution of the population imbalance with different nonlinear tunneling parameters.  Initially, assuming that $(\phi(0),z(0))=(0,-0.6)$, it is obvious that the system presents a Josephson oscillation evolution, which has stable amplitude. However, as the nonlinear tunneling effect increasing, the period of the oscillation becomes smaller. Therefore, one can come to a conclusion that the nonlinear tunneling can not influence the population of the atoms in double well, but will speed up the tunneling of atoms. In other words, we can control the experimentally observed rate of atoms in double well system by changing the nonlinear tunneling coupling strength to some extent.
\begin{figure}[H]
\centering
\includegraphics[width=3.5in]{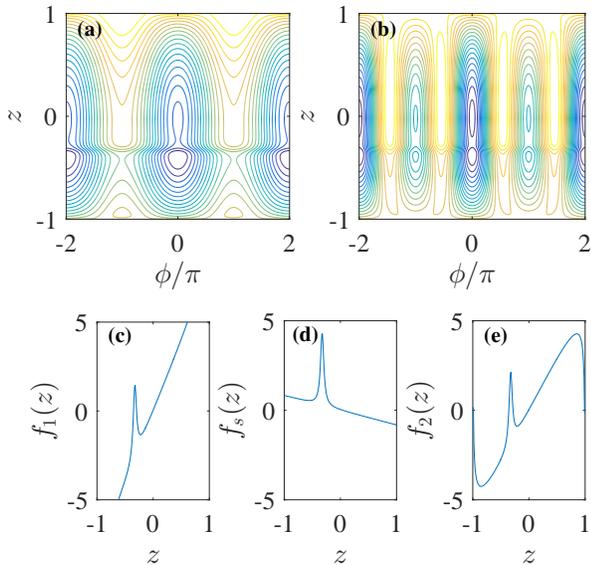}
\caption{(Color online) Energy contours of a Bose Josephson junction, with the nonlinear tunneling term $(a)$ $S=0.1$, $(b)$ $S=3.87$. (c), (d) and (e) Gradient of the energy along the line $\phi=0$, $\phi=\frac{\pi}{2}$ and $\phi=\pi$, respectively. The other parameters are $NV/(2\Omega)=r=3$, $NV^{\prime}/(2\Omega)=r_{bc}=0.1$, $\widetilde{A}=0.02$, $B=-0.65$, and $C=0.07$.}
\end{figure}
\begin{figure}[H]
\centering
\includegraphics[width=3.5in]{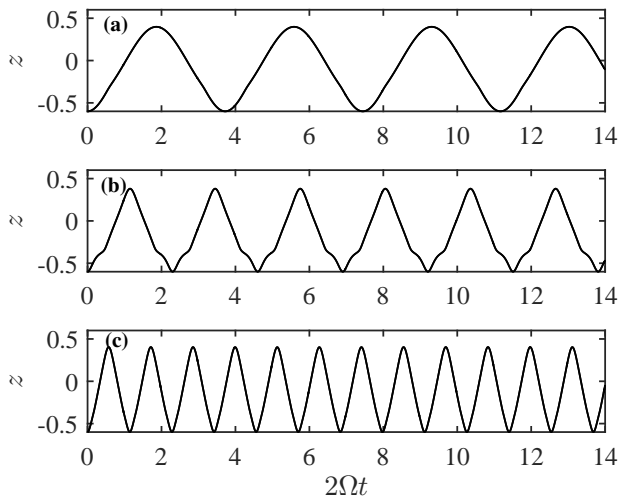}
\caption{The time evolution of the population imbalance[in units of $2\Omega/(\delta U_0)$] versus the reduced time $2\Omega t$. The initial conditions are $(\phi(0),z(0))=(0,-0.6)$, $r_{bc}=3$, $r_b(r_c)=0.1$ with different $S$ with $(a)$ $S=0$, $(b)$ $S=1$, and $(c)$ $S=2$. The parameters are same as ones in Fig. $2$.}
\end{figure}
\begin{figure}[H]
\centering
\includegraphics[width=3.5in]{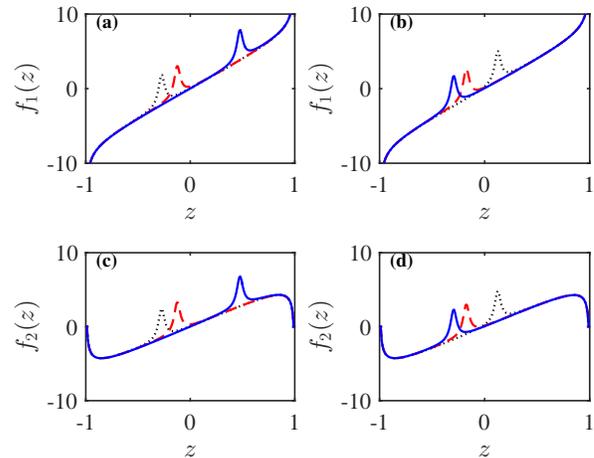}
\caption{The gradient of the energy along the line $\phi=0$ and $\phi=\pi$, respectively. The left-hand column shows changes in the location and number of fixed points with parameters $E=0.1$ and $D=0.1$ (black dotted lines), $D=0.2$ (red dashed lines), $D=0.4$ (blue solid lines), along the lines $\phi=0$ in Fig. $4(a)$ and $\phi=\pi$ in Fig. $4(c)$. The right-hand column shows changes in the location and number of fixed points with parameter $D=0.3$ and $E=0.1$ (black dotted lines), $E=0.3$ (red dashed lines), $E=1.5$ (blue solid lines), along the lines $\phi=0$ of Fig. $4(b)$ and $\phi=\pi$ of Fig. $4(d)$. $S=3.87$. The same parameters as in Fig. $2$.}
\end{figure}
\textit{With the moving end mirror}. In order to compare the difference of the dynamical of BECs DW  between the moving end mirror and without the moving end mirror, we assume $D\not=0$, $E\not=\omega_{p}$. First of all, we solved the roots of Eq. $(3.7)$ with different parameters numerically, as shown in Figs. $4(a)$-$4(d)$. Fig. $4(a)$ indicates that the stability point of system can move upwards with the increasing coupling strength between the cavity and the mirror along the lines $\phi=0$ when the detuning between the pump field and the mirror has fixed value. However, while the coupling strength between the cavity and the mirror has a fixed value, an opposite result appears with detuning increasing, as shown in Fig. $4(b)$. So, in the specific sets of parameters, we have the flexibility to control the number of stable points. From Figs. $4(c)$ and $4(d)$, we can obtain the same results along the lines $\phi=\pi$. As is well known, the loss of stability of a semiclassical stability points is associated with an entanglement in the steady state of the full quantum system, the semiclassical dynamics of the system undergoes a bifurcation of the stability point corresponding to the quantum steady state, and the maximum entanglement occurs at the parameter values about bifurcation of the stability point in a dissipative many-body system\cite{Schneider2002Entanglement}. So, we can conjecture that adjusting the coupled strength and detuning between the moving end mirror and the pump field can well control the entanglement in the steady state of the full quantum system.

\begin{figure*}
\centering
\includegraphics[width=7in]{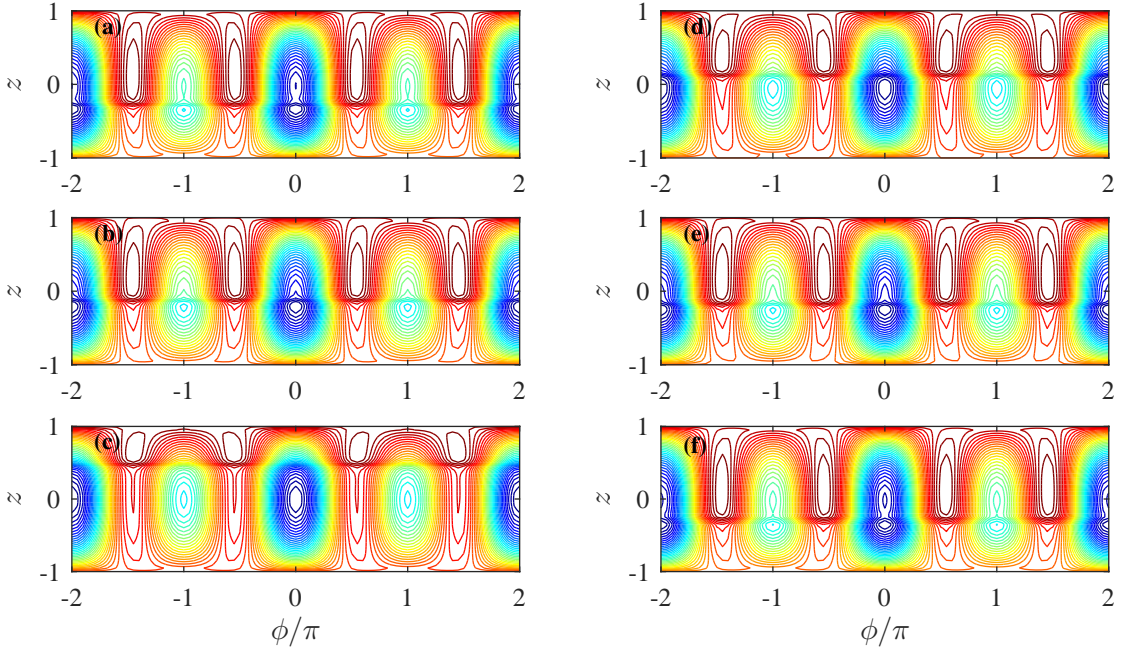}
\caption{Energy contours of a Bose Josephson junction. The left-hand column shows changes of the phase diagram  with $E=0.1$ and different parameter $D$ in (a) $D=0.1$, (b) $D=0.2$, and (c) $D=0.4$. The right-hand column shows changes of the phase diagram  with $D = 0.3$ and different $E$  in $(d)$ $E = 0.1$ , $(e)$ $E = 0.3$, and $(f)$ $E = 1.5$. The same parameters as in Fig.$ 4$.}
\end{figure*}
\begin{figure}[H]
\centering
\includegraphics[width=3.5in]{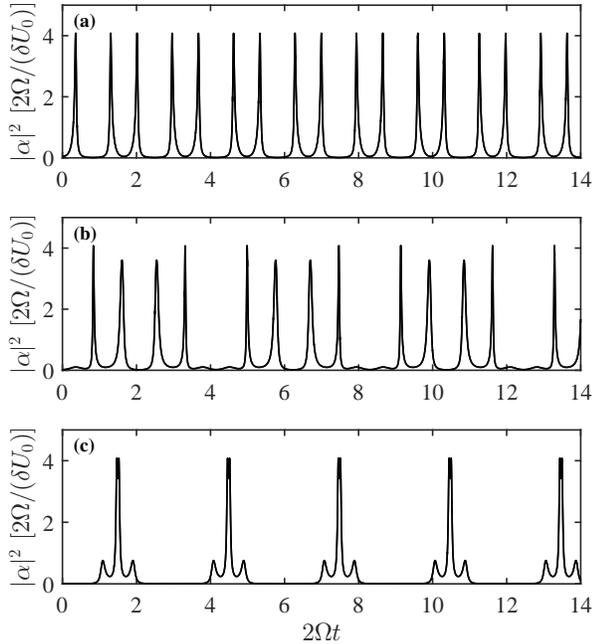}
\caption{Intra cavity photon number $|\alpha|^2$ [in units of $2\Omega/(\delta U_0)$] versus the reduced time $2\Omega t$ with different parameters $(a)$ $D=0.1$, $(b)$ $D=0.2$ and $(c)$ $D=0.4$. The initial conditions are $(\phi(0),z(0))=(0,-0.6)$, $E=0.1$, $S=1.37$. The same parameters as in Fig. $2$.}
\end{figure}

As the stability point of system has a great relationship with the phase diagram, we further plot the phase diagram with different coupling strengths and detunings between the cavity field and the mirror to discuss the occurrence of the bifurcation of the stability point. The left-hand column shows the changes of the phase diagram with the increasing coupling strength between the cavity and the mirror for a fixed detuning in Fig. $5$. There are three stability points for $\phi=0$, one stability point is localized near by $\phi=\pi/2$, and three stability points for $\phi=\pi$ for the small coupling strength between the cavity and the moving end mirror, as shown in Fig. $5(a)$. The typical character of the level curves has undergone tremendous changes with the coupling strength increasing. For example, in Fig. $5(c)$, there are one stability point for $\phi=0$, one stability point is localized near by $\phi=\pi/2$, and one stability points for $\phi=\pi$ while the coupling strength between the cavity and the moving end mirror is strong. This means that the system experiences a stability point bifurcations for certain choices of the coupling parameters. From Eqs. $(3.5)$ , $(3.7)$ and Fig. $4$, we can see that the distribution of the phase diagram is symmetric about $z = 0$ without the cavity-field induced tilt. However, this symmetry will be broken when the coupling strength between the atoms and the cavity when the moving end mirror exists. In this case, there are three stability points along the lines $\phi=0$ or $\phi=\pi$ in this system, which satisfies $-B+D^2/E<0$. And only one stability point exists when $-B+D^2/E>0$. In addition, we can obtain that the absolute value of $-B+D^2/E$ represents the distance of two maximum(two minimum) in phase diagram from Figs. $4$ and $5$. One of the maxima (minimum) is localized in $z = 0$ and another location of the maximum (minimum) depends on the absolute value of $-B+D^2/E$. Therefore, when adding the moving end mirror, we can regulate the values of $D$ and $E$ to control the distance between two maxima (minimum) in the phase diagram and the number of stability points of the system. The right-hand column shows changes of the phase diagram with the increasing detunings between the pump field and the mirror when the coupling strength has a fixed value in Fig. $5$. Obviously, one will get a opposite variation tendency compared with the result of the left-hand column presenting. Then, the semiclassical dynamics of the system undergoes a bifurcation of the stability points with the coupling and the detuning, but the behavior is different.

\begin{figure}[H]
\centering
\includegraphics[width=3in]{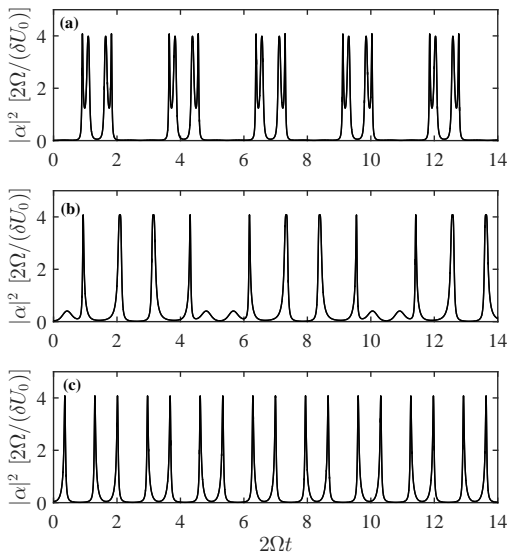}
\caption{Intra cavity photon number $|\alpha|^2$ [in units of $2\Omega/(\delta U_0)$] versus the reduced time $2\Omega t$ with different parameters $(a)$ $E=0.1$, $(b)$ $E=0.4$ and $(c)$ $E=0.9$. The initial conditions are $(\phi(0), z(0))=(0,-0.6)$, $D=0.3$, $S=1.37$. The same parameters as in Fig. $2$.}
\end{figure}

The outputs of the cavity mode carries a lot of information about the population of the atoms
between the two traps as it leaks out of the cavity. From Eqs. $(3.3)$-$(3.5)$, we note that the distribution of photon numbers is influenced by two factors. On the one hand, the initial conditions of system evolution. On the other hand, the evolution of the conjugate variables $z$ and $\phi$ of the energy determined by the initial conditions. The different energy curves correspond to different distributions of photon numbers. For example, when the system is in a stability state, the distributions of photon numbers is constant. When the state of the system evolves along the energy curves around the stability point, the distribution of photon numbers has a small changes with a period. When the state of the system evolves is a oscillation over the range of the population imbalance, the distribution of photon numbers has enormous changes. Comparing with no moving end mirror, the coupling between the cavity and the moving end mirror and the detuning between the pump field and the moving end mirror make the energy curve change, the distribution of photon numbers are also changed. As shown in Figs. $6$ and $7$, we plot the number of intra cavity photons with different coupling strengths between the pump field and the moving end mirror, and different detunings between the cavity and the moving end mirror respectively. It can be found that although the variety of detunings and coupling strengths are small, the change of the difference in the outputs of the cavity have enormous difference. Comparing with Ref.\cite{zhang2008mean}, Lorenzian appears six peaks in Figs. $6$ and $7$ , as can be seen from equations Eqs. $(3.4)$ and $(3.5)$, which is due to the nonlinear tunneling term. In addition, the coupling strengths and the detunings between the cavity and mirror only shift the center of the photon number distribution, and do not change the distribution of photon number.

\section{Conclusion}
In this paper, we have investigated two component BECs DW in the optomechanical cavity with a nonlinear tunneling interaction and the moving end mirror. We used the mean-field method to obtain the dynamical equation of BECs DW based on the two- mode approximation and found that the model exhibited abundant dynamical information of the BECs DW. The introduction of the nonlinear tunneling leads to an increase of the system stability points along near by $\pi/2$, and the distribution of photon numbers is very different from ones without the nonlinear tunneling term, which makes the phase diagram of the system more become rich. In addition, as the nonlinear tunneling interaction strength increases, the distribution period of the number of particles becomes smaller and smaller. It is clear that the moving end mirror has little effect on the population of the atoms between the two traps and the phase diagram of the system, but the coupling strength between the cavity and the moving end mirror and the detuning between the pump field and the moving end mirror, as the degree of freedom of the system can regulate the number of the stability points of the system, and then can control the dynamics of the one. We can control a bifurcation of the stability points of the system by changing the parameters of the mirror-cavity interaction. This is also very helpful to the study of the entanglement of the system.

\begin{acknowledgments}
This work was supported by NSFC under grants Nos. 11274148 and  11434015, the National Key R$\&$D Program of China under grants Nos. 2016YFA0301500, and SPRPCAS under grants No. XDB01020300, XDB21030300.
\end{acknowledgments}

\end{document}